\documentstyle[aps,prl,preprint,floats,epsfig]{revtex}

\begin{document}

\preprint{\tighten\vbox{\hbox{\hfil CLNS 00/1667}
                        \hbox{\hfil CLEO 00-5}
}}

\title{\LARGE Measurements of Charm Fragmentation into $D_s^{*+}$ and $D_s^+$ in $e^+e^-$ Annihilations
 at $\sqrt{s}=10.5$ GeV }

\author{CLEO Collaboration}
\date{\today}

\maketitle
\tighten

\begin{abstract} 
A study of charm fragmentation into $D_s^{*+}$ and $D_s^+$ in $e^+e^-$ 
annihilations at $\sqrt{s}=10.5$ GeV is presented. This study using 
$4.72\pm0.05$ fb$^{-1}$ of CLEO II data reports measurements of the 
cross-sections $\sigma(D_s^{*+})$ and $\sigma(D_s^+)$ in momentum regions 
above $x=0.44$, where $x$ is the $D_s$ momentum divided by the maximum 
kinematically allowed $D_s$ momentum. The $D_s$ vector to vector plus 
pseudoscalar production ratio is measured to be 
$P_V(x(D_s^+)>0.44) = 0.44 \pm 0.04$. 
\end{abstract}
\newpage

{
\renewcommand{\thefootnote}{\fnsymbol{footnote}}

\begin{center}
R.~A.~Briere,$^{1}$
B.~H.~Behrens,$^{2}$ W.~T.~Ford,$^{2}$ A.~Gritsan,$^{2}$
H.~Krieg,$^{2}$ J.~Roy,$^{2}$ J.~G.~Smith,$^{2}$
J.~P.~Alexander,$^{3}$ R.~Baker,$^{3}$ C.~Bebek,$^{3}$
% ...
\end{center}
 
\small
\begin{center}
$^{1}${Carnegie Mellon University, Pittsburgh, Pennsylvania 15213}\\
$^{2}${University of Colorado, Boulder, Colorado 80309-0390}\\
$^{3}${Cornell University, Ithaca, New York 14853}\\
% ...
\end{center}

\setcounter{footnote}{0}
}
\newpage


\section{Introduction}

The production cross-sections of $q{\bar q}$ pairs in $e^+e^-$ annihilations 
can be calculated using QCD, but the process of fragmentation whereby hadrons 
are formed is non-perturbative and phenomenological models are used to 
describe it. Two properties of hadron production that can be experimentally 
measured are the hadron momentum distribution and the relative population of 
available spin states.

Measurements of primary hadron fragmentation can be challenging due to 
cascades from higher order resonances that can be indistinguishable from the 
primary hadrons. The study of $D_s^+$ and $D_s^{*+}$ fragmentation
in $e^+e^-$ annihilations at $\sqrt{s}=10.5$ GeV benefits from the fact that 
$L=1$ charm mesons have not been observed to decay to either $D_s^+$ or 
$D_s^{*+}$~\cite{pdg} and the influence of $B$ events is kinematically 
eliminated for $x(D_s)>0.4$, where $x$ is the $D_s$ momentum divided by the 
maximum kinematically allowed $D_s$ momentum. The $D_s$ system is thus 
particularly well suited for the measurement of the vector to pseudoscalar 
production ratio. 

The vector to pseudoscalar production ratio is usually described using the 
variable
\begin{equation}
P_V = \frac{V}{V+P}~, \label{eq:pv1}
\end{equation}
where P and V represent, respectively, the number of pseudoscalar and vector 
mesons directly produced through a particular production mechanism, e.g. 
$e^+e^-$ annihilations. Counting the number of spin states available to an 
$L=0$ meson leads to the expectation that $P_V=0.75$. This spin counting model
has been shown to be useful for describing the $D^{*+}$ spin 
alignment~\cite{align}, but most measured values of $P_V$ have been 
significantly lower than 0.75 for charm mesons. Other models based upon the 
mass difference between the vector and pseudoscalar states predict values of 
$P_V$ that are less than 0.75~\cite{pvmodel}, but more precise measurements 
are needed to better determine any relationship between $P_V$ and the 
mass difference.


\section{Detector and Event Selection}

The data in this analysis were collected from $e^+e^-$ collisions at the 
Cornell Electron Storage Ring (CESR) by the CLEO II detector. The CLEO II 
detector is a general purpose charged and neutral particle spectrometer 
described in detail elsewhere~\cite{nim}. The dataset used in this analysis 
contains $3.11\pm0.03$ fb$^{-1}$ of data collected at the $\Upsilon(4S)$ 
resonance and $1.61\pm0.02$ fb$^{-1}$ of data collected below the $b{\bar b}$ 
threshold (about 60 MeV below the $\Upsilon(4S)$ resonance), for an 
approximate total of $5 \times 10^6$ $c{\bar c}$ events. 

In this analysis, $D_s^{*+}$ mesons are reconstructed via the decay 
$D_s^{*+} \to D_s^+ \gamma$ and $D_s^+$ mesons are reconstructed via the decay
chain $D_s^+ \to \phi \pi^+$ with $\phi \to K^+ K^-$ (inclusion of charge 
conjugate modes is implied throughout this paper). 

All charged tracks used in this analysis are required to have an origin close 
to the $e^+e^-$ interaction region and must be well reconstructed. When drift 
chamber particle identification information is available, the specific 
ionization, $dE/dx$, must be within two standard deviations of the expected 
value for candidate kaon tracks and within three standard deviations of the 
expected value for candidate pion tracks. 

Showers in the crystal calorimeter are considered as photon candidates if 
they have a minimum energy of 100 MeV, are within either the barrel 
($|\cos \theta_s|<0.71$, where $\theta_s$ is the angle between the shower 
and the $e^+$ beam direction) or endcap ($0.85<|\cos \theta_s|<0.95$) regions,
have an energy deposition consistent with that expected for a photon, and do 
not include any crystals near a projected charged track.

Candidate $\phi$ mesons are reconstructed using all appropriately signed 
combinations of candidate kaon tracks in an event. The invariant mass $M(KK)$ 
is required to be within 8.4 MeV/$c^2$ (approximately 2 standard deviations) 
of the known $\phi$ mass~\cite{pdg}. Candidate $D_s^+$ mesons are 
reconstructed using all combinations of $\phi$ candidates and candidate pion 
tracks in an event. Candidate $D_s^{*+}$ mesons are reconstructed using
candidate photons, and $\phi\pi$ combinations with invariant mass $M(KK\pi)$ 
within 20 MeV/$c^2$ (approximately 2.5 to 3 standard deviations) of the known
$D_s^+$ mass.

Because the $\phi$ must be polarized in the helicity-zero state in a 
$D_s \to \phi\pi$ decay, the decay of the $\phi$ has an angular
distribution proportional to $cos^2 \alpha$, where $\alpha$ is the angle 
between the $K^+$ and $D_s^+$ momentum vectors in the $\phi$ rest frame. 
Since the background angular distribution is flat, the signal to background 
ratio is improved by requiring $|cos \alpha| >0.35$. The signal to background 
ratio is further enhanced by requiring that $cos \theta_\pi \geq -0.8$, where 
$\theta_\pi$ is the angle of the $\pi$ momentum vector in the $D_s^+$ rest 
frame relative to the $D_s^+$ momentum vector in the laboratory frame; the 
signal distribution is flat in this variable while background events peak at 
$\cos \theta_\pi =-1.0$. Because of the minimum energy restriction for photon
candidates, signal photons traveling in a direction opposite to the 
$D_s^{*+}$ direction in the laboratory frame are excluded from the candidate 
sample. By requiring $\cos \theta_\gamma >-0.8$, where $\theta_\gamma$ is 
defined as the angle of the photon momentum vector in the $D_s^{*+}$ rest 
frame relative to the $D_s^{*+}$ momentum vector in the laboratory frame, 
additional background $D_s^{*+}$ candidates are suppressed.

Low momentum $D_s^+$ candidates are difficult to analyze because of the 
large amount of background from combinatorics as well as $B$ decays. 
The analysis is therefore restricted to $x(D_s^+)>0.44$ where 
\begin{equation}
x(D_s^+)\equiv\frac{p(D_s^+)}{p_{max}(D_s^+)}~, \label{eq:xds}
\end{equation}
and 
\begin{equation}
p_{max}(D_s^+)=\sqrt{E_{beam}^2-m_{D_s^+}^2}~. \label{eq:pmaxds}
\end{equation}
For $D_s^{*+}$ candidates, the $x(D_s^+)$ requirement is replaced by 
$x(D_s^{*+})>0.5$ where
\begin{equation}
x(D_s^{*+})\equiv\frac{p(D_s^{*+})}{p_{max}(D_s^{*+})}~,\label{eq:xdss}
\end{equation}
and
\begin{equation}
p_{max}(D_s^{*+})=\sqrt{\left(E_{beam}-\frac{m^2_{D_s^{*+}}-m^2_{D_s^+}}
{4 E_{beam}}\right)^2-m^2_{D_s^+}}~.\label{eq:pmaxdss}
\end{equation}
In principle, $B\to D_s^+ \pi$ can result in $x(D_s^+) \sim 0.5$. However, 
such decays are $b\to u$ transitions and thus heavily suppressed, so they are 
expected to be a negligible source of background.

Based on the assumption that all observed $D_s^{*+}$ are primary, the 
$D_s^{*+}$ momentum spectrum is simply studied by measuring the $D_s^{*+}$ 
yield in eight equal sized bins of $x(D_s^{*+})$ over the range 
$0.5<x(D_s^{*+})<0.98$. However, the observed $D_s^+$ can be primary or 
$D_s^{*+}$ daughters. In order to study the momentum distribution of primary 
$D_s^+$ mesons, it is necessary to subtract out the $D_s^{*+}$ 
contribution to the $D_s^+$ yields. Since all $D_s^{*+}$ are assumed to 
decay to $D_s^+$, the $D_s^+$ yields from $D_s^{*+}$ decays can be accounted 
for by simply measuring the $D_s^{*+}$ yields as above, but in bins of the 
variable $x(D_s^+)$ rather than $x(D_s^{*+})$. After the $D_s^{*+}$ yields 
are corrected for efficiency and the branching ratio 
${\cal B}(D_s^{*+}\to D_s^+\gamma)$, they are subtracted from the efficiency 
corrected $D_s^+$ yield in each $x(D_s^+)$ bin to calculate the primary 
$D_s^+$ yield.


\section{Fitting}

The $D_s^{*+}$ yields are projected onto $\Delta M = M(KK\pi\gamma)-M(KK\pi)$ 
for $D_s^{*+}$ candidates and the $D_s^+$ yields are projected onto $M(KK\pi)$
for $D_s^+$ candidates. Fitting shapes for the peaks in these distributions 
are determined using a sample of Monte Carlo events generated using the Lund 
\textsc{jetset} 7.3~\cite{jetset} program combined with a 
\textsc{geant}-based CLEO II detector simulation, where every event contains 
a $D_s^{*+}$ or $D_s^+$ decaying through the modes specified above. 

\begin{figure}
\centerline{\psfig{file=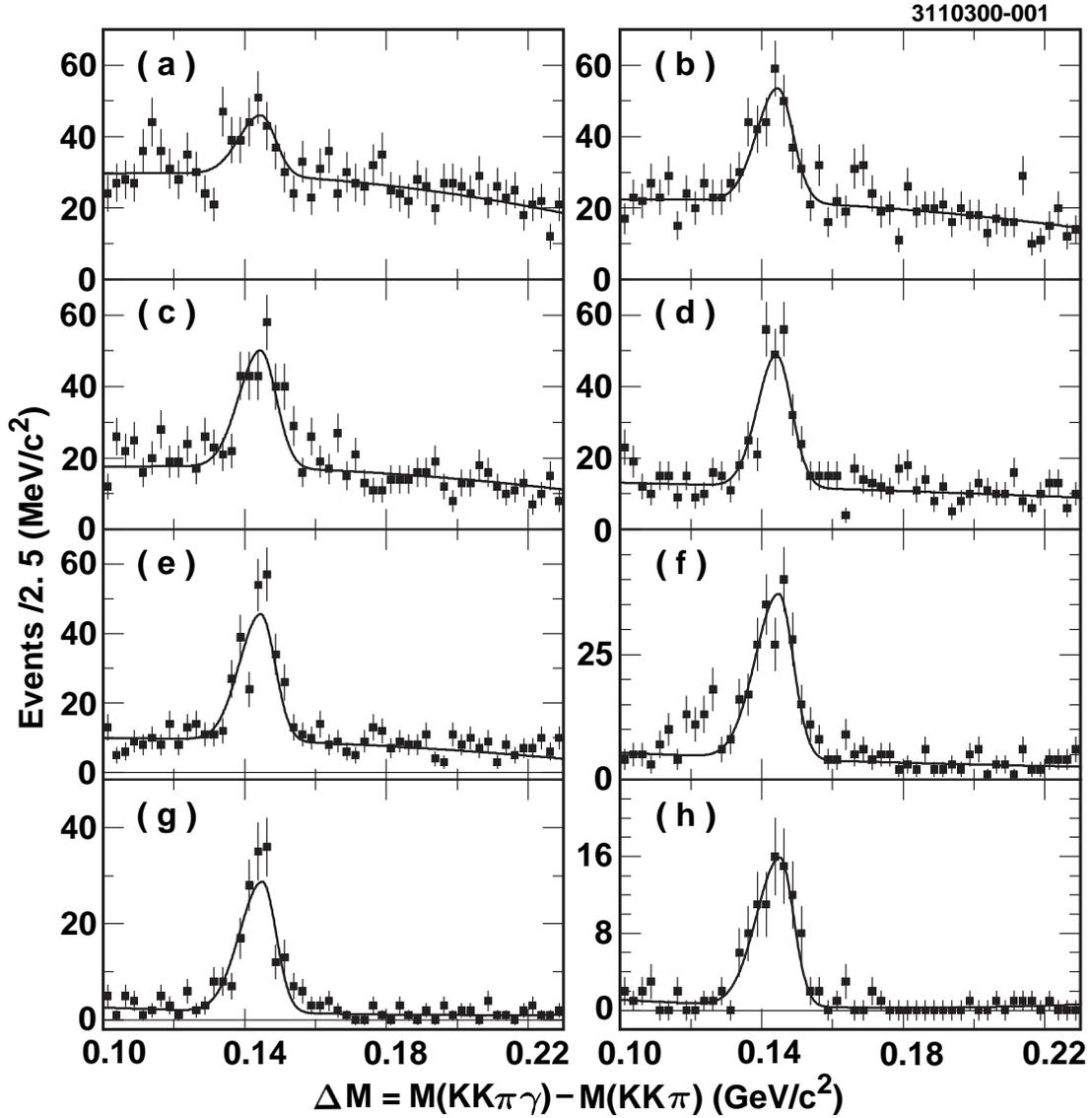,height=6in}}
\caption{{\label{fig:dssplot}}
Fits to the $\Delta M=M(KK\pi\gamma)-M(KK\pi)$ distributions for candidate 
$D_s^{*+}$ events that are used to determine the $D_s^{*+}$ yields in the 
eight $x(D_s^{*+})$ ranges
(a) $0.50 - 0.56$, (b) $0.56 - 0.62$, (c) $0.62 - 0.68$,
(d) $0.68 - 0.74$, (e) $0.74 - 0.80$, (f) $0.80 - 0.86$, 
(g) $0.86 - 0.92$, and (h) $0.92 - 0.98$.
}
\end{figure}

\begin{figure}
\centerline{\psfig{file=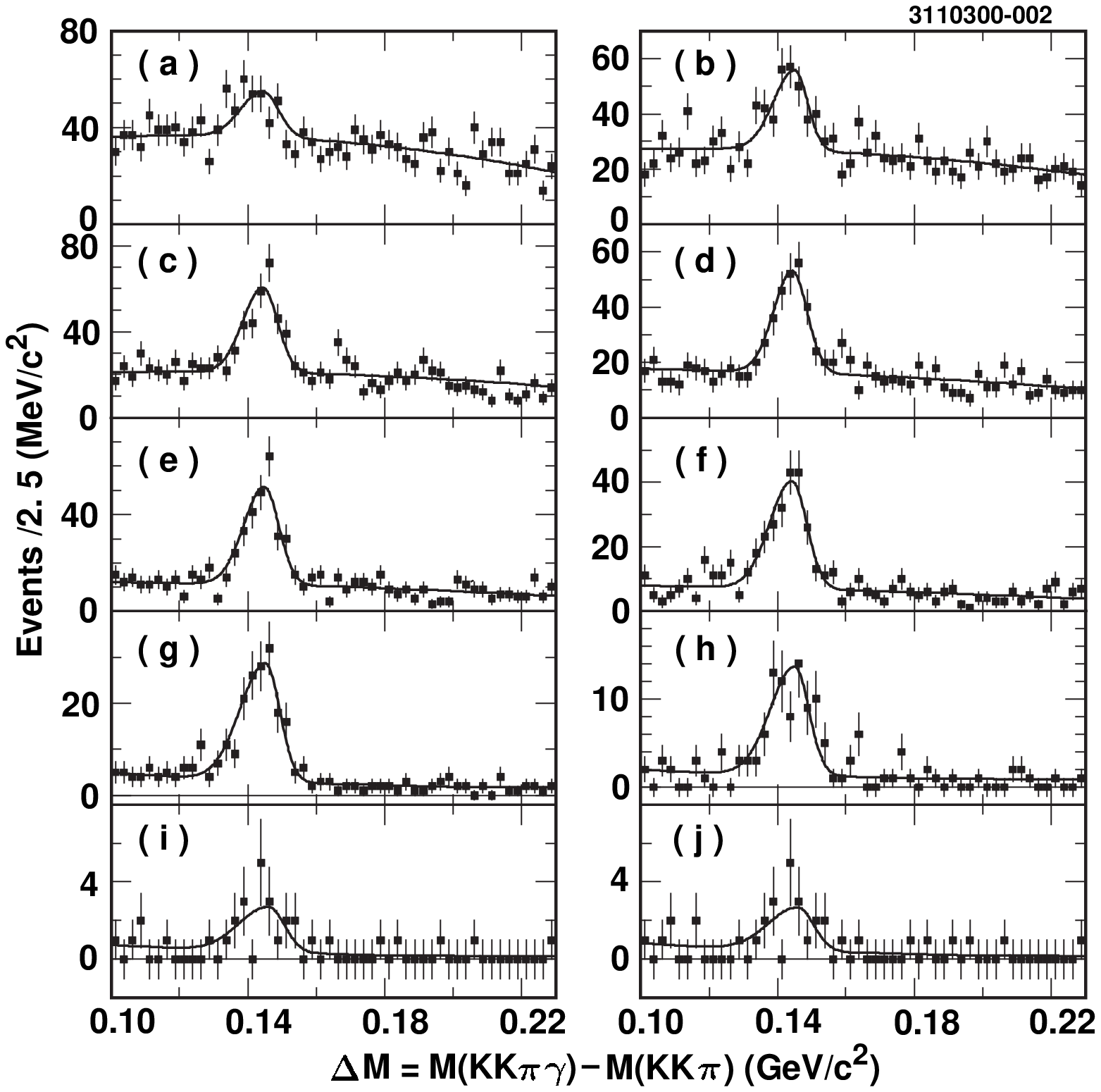,height=6in}}
\caption{{\label{fig:dsssubplot}}
Fits to the $\Delta M=M(KK\pi\gamma)-M(KK\pi)$ distributions for candidate 
$D_s^{*+}$ events that are used to determine the $D_s^{*+}$ yields in the 
ten $x(D_s^+)$ ranges
(a) $0.44 - 0.50$, (b) $0.50 - 0.56$, (c) $0.56 - 0.62$, (d) $0.62 - 0.68$,
(e) $0.68 - 0.74$, (f) $0.74 - 0.80$, (g) $0.80 - 0.86$, 
(h) $0.86 - 0.92$, (i) $0.92 - 0.98$, and (j) $0.92 - 1.00$.
}
\end{figure}

The $\Delta M$ distributions in data and the signal Monte Carlo sample are 
simultaneously fit to the sum of an asymmetric Gaussian for the signal and 
separate second-order Chebyshev polynomials for the background in each 
distribution. An asymmetric Gaussian is used because of the larger tail on the
lower side of the peak attributable to energy leakage in the calorimeter. 
The fits to data used to determine the $D_s^{*+}$ yields in the selected 
regions of $x(D_s^{*+})$ and $x(D_s^+)$ are shown in Figs. \ref{fig:dssplot} 
and \ref{fig:dsssubplot}. 

\begin{figure}
\centerline{\psfig{file=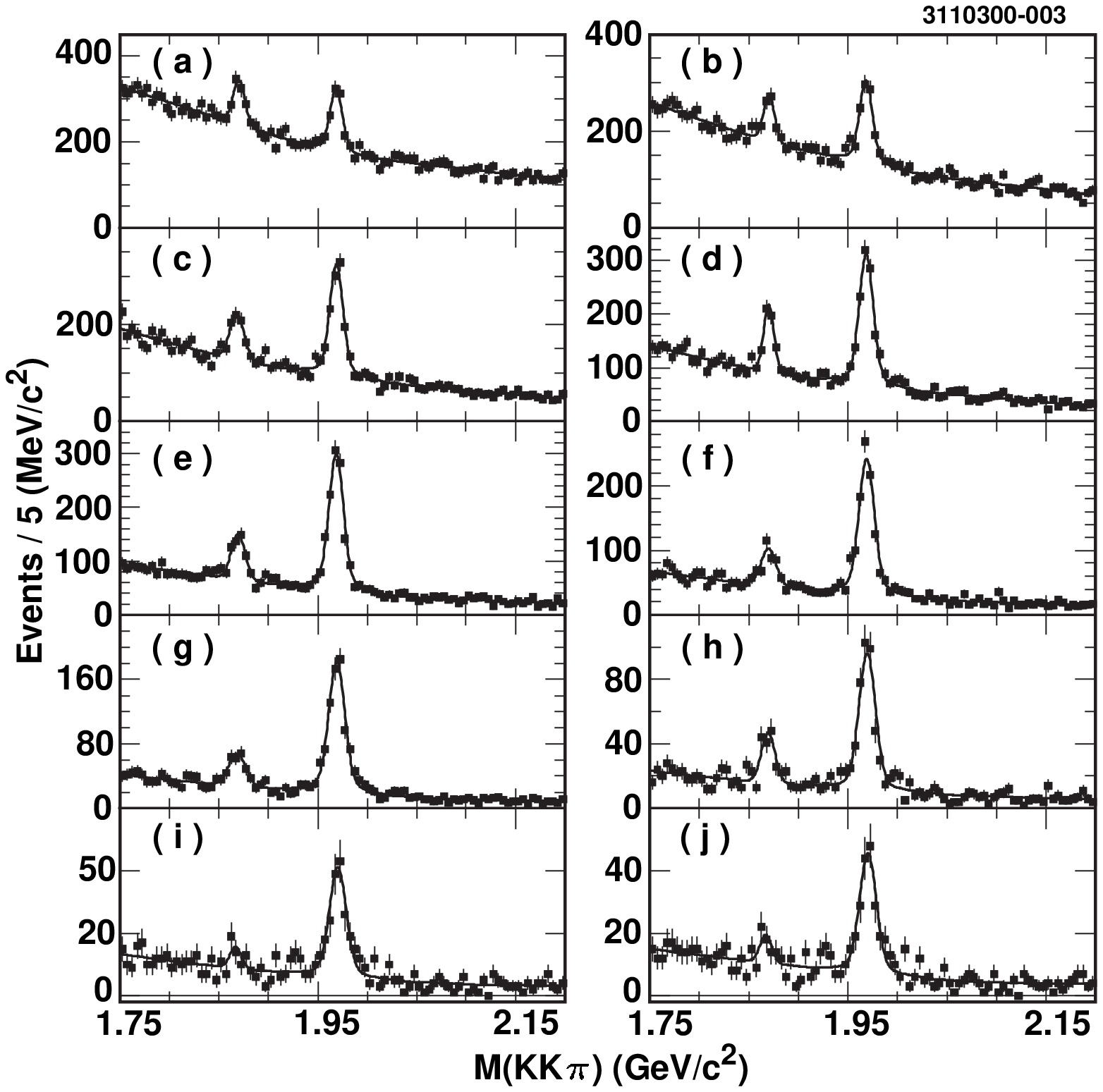,height=6in}}
\caption{{\label{fig:dsplot}}
Fits to the $M(KK\pi)$ distributions for candidate $D_s^+$ events 
that are used to determine the $D_s^+$ yields in the ten $x(D_s^+)$ ranges
(a) $0.44 - 0.50$, (b) $0.50 - 0.56$, (c) $0.56 - 0.62$, (d) $0.62 - 0.68$,
(e) $0.68 - 0.74$, (f) $0.74 - 0.80$, (g) $0.80 - 0.86$, 
(h) $0.86 - 0.92$, (i) $0.92 - 0.98$, and (j) $0.92 - 1.00$.
}
\end{figure}

The $M(KK\pi)$ distributions in data and the signal Monte Carlo sample are 
simultaneously fit to the sum of a double Gaussian with common mean for the 
$D_s^+$ signal, a Gaussian for the $D^+$ signal, two straight lines joined by 
a quadratic for the combinatoric background in data, and a first order 
Chebyshev polynomial for the small amount of background in the Monte Carlo 
sample. The fits to data used to determine the $D_s^+$ yields in the selected
regions of $x(D_s^+)$ are shown in Fig. \ref{fig:dsplot}.

\section{Efficiencies}
The $D_s^+$ and $D_s^{*+}$ detection efficiencies are estimated using a 
sample of Monte Carlo events that contains signal as well as background events
and is independent of the signal Monte Carlo sample used in the fitting 
procedure. The $D_s^{*+}$ efficiency values in the $x(D_s^{*+})$ regions are 
listed in Table {\ref{tab:dsseff}}, while the $D_s^+$ and $D_s^{*+}$ 
efficiencies in the $x(D_s^+)$ regions are listed in Table \ref{tab:dseff}.

\begin{table}
\centering
\begin{tabular}{ c c }
$x(D_s^{*+})$ region & $D_s^{*+}$ Efficiency \\ \hline
0.50 - 0.56 & $0.172 \pm 0.008$ \\
0.56 - 0.62 & $0.175 \pm 0.006$ \\
0.62 - 0.68 & $0.178 \pm 0.005$ \\
0.68 - 0.74 & $0.181 \pm 0.004$ \\
0.74 - 0.80 & $0.184 \pm 0.004$ \\
0.80 - 0.86 & $0.187 \pm 0.005$ \\
0.86 - 0.92 & $0.190 \pm 0.006$ \\
0.92 - 0.98 & $0.193 \pm 0.008$ \\
(0.92 - 1.00) & ($0.194 \pm 0.008$) 
\end{tabular}
\caption{{\label{tab:dsseff}} 
$D_s^{*+}$ detection efficiencies in the specified regions of $x(D_s^{*+})$, 
where the efficiency has been smoothed using a fit to the raw efficiency 
spectrum with a first order Chebyshev polynomial. The values that are in 
parentheses are used only for the calculation of the total $D_s^{*+}$ yield 
and cross-section for $x(D_s^{*+})>0.5$.
}
\end{table}

\begin{table}
\centering
\begin{tabular}{ c c c }
$x(D_s^+)$ region & $D_s^+$ Efficiency & $D_s^{*+}$ Efficiency\\ \hline
0.44 - 0.50 & $0.366 \pm 0.010$ & $0.180 \pm 0.009$\\
0.50 - 0.56 & $0.369 \pm 0.008$ & $0.179 \pm 0.007$\\
0.56 - 0.62 & $0.373 \pm 0.007$ & $0.178 \pm 0.006$\\
0.62 - 0.68 & $0.376 \pm 0.005$ & $0.177 \pm 0.005$\\
0.68 - 0.74 & $0.379 \pm 0.005$ & $0.176 \pm 0.004$\\
0.74 - 0.80 & $0.382 \pm 0.005$ & $0.175 \pm 0.005$\\
0.80 - 0.86 & $0.386 \pm 0.007$ & $0.174 \pm 0.006$\\
0.86 - 0.92 & $0.389 \pm 0.008$ & $ 0.140 \pm 0.027$ \\
0.92 - 0.98  & $0.392 \pm 0.010$ & $ 0.248 \pm 0.107$ \\
(0.92 - 1.00) & ($0.393 \pm 0.011$) & ($0.248 \pm 0.106$)
\end{tabular}
\caption{{\label{tab:dseff}}
$D_s^+$ and $D_s^{*+}$ detection efficiencies in the specified regions 
of $x(D_s)$ where the efficiency has been smoothed using a fit to the raw 
efficiency spectrum with a first order Chebyshev polynomial. The values that 
are in parentheses are used for the calculation of $P_V(D_s)$. The $D_s^{*+}$ 
efficiency values with $x(D_s^+)>0.86$ are excluded from the smoothing 
process because they are not expected to be modeled by the same function used 
for $x(D_s^+)<0.86$.
}
\end{table}

For the $D_s^{*+}$ production study, the efficiency for each $x(D_s^{*+})$ bin
is measured using the fitting procedure described above. The binned raw 
efficiency values within the range $0.50<x(D_s^{*+})<0.98$ are fit with a 
first order Chebyshev polynomial to provide a smoothly varying efficiency as 
a function of $x(D_s^{*+})$. The smoothed efficiency value at the center of 
each $x(D_s^{*+})$ region is used to calculate the efficiency corrected 
$D_s^{*+}$ yield and cross-section.

For the $D_s^+$ fragmentation study, the $D_s^+$ and $D_s^{*+}$  efficiencies 
are measured in each $x(D_s^+)$ bin using the fitting procedure described 
above. The binned raw $D_s^+$ efficiencies within the range 
$0.44<x(D_s^+)<0.98$ are fit with a first-order Chebyshev and the smoothed 
efficiency values are used to calculate the efficiency corrected $D_s^+$ 
yields. The binned raw $D_s^{*+}$ efficiencies with $0.44<x(D_s^+)<0.86$ are 
fit with a first-order Chebyshev polynomial but the efficiencies in the 
region $x(D_s^+)>0.86$ are excluded from the fit because of expected 
efficiency loss due to the larger proportion of photons in that region with 
energies less than 100 MeV. The smoothed efficiency values are used to 
calculate the efficiency corrected $D_s^{*+}$ yields for $x(D_s^+)<0.86$, 
while the raw efficiency values are used for $x(D_s^+)>0.86$. 


\section{Results}
The $D_s^{*+}$ yields, efficiency corrected yields and cross-sections in the 
eight $x(D_s^{*+})$ regions are all listed in Table {\ref{tab:xdssyield}. 
These same quantities for $D_s^+$ and $D_s^{*+}$ in the nine $x(D_s^+)$ 
regions are listed in Tables \ref{tab:xdsyield1} and \ref{tab:xdsyield2},
respectively. The calculated primary $D_s^+$ yields and cross-sections are 
presented in Table \ref{tab:dsprim}. 

\begin{table}
\centering
\begin{tabular}{ c c c c }
                     & Measured         & Efficiency Corrected & \\
$x(D_s^{*+})$ region & $D_s^{*+}$ Yield & $D_s^{*+}$ Yield & 
${\cal B}\cdot\sigma(D_s^{*+})$(pb) \\ \hline
0.50 - 0.56 & $84 \pm 18$     & $488 \pm 116$   & $0.10 \pm 0.02$\\
0.56 - 0.62 & $159 \pm 19$    & $911 \pm 138$   & $0.19 \pm 0.03$\\
0.62 - 0.68 & $173 \pm 19$    & $972 \pm 137$   & $0.21 \pm 0.03$\\
0.68 - 0.74 & $183 \pm 17$    & $1014 \pm 131$  & $0.21 \pm 0.03$\\
0.74 - 0.80 & $202 \pm 17$    & $1095 \pm 135$  & $0.23 \pm 0.03$\\
0.80 - 0.86 & $172 \pm 16$    & $918 \pm 117$   & $0.19 \pm 0.02$\\
0.86 - 0.92 & $148 \pm 13$    & $778 \pm 100$    & $0.16 \pm 0.02$\\
0.92 - 0.98 & $84 \pm 10$     & $434 \pm 65$    & $0.09 \pm 0.01$\\
(0.92 - 1.00) & ($99 \pm 11$) & ($511 \pm 74$)  & ($0.11 \pm 0.02$) \\ \hline
0.50 - 1.00 & $1219 \pm 47$ & $6687 \pm 260 \pm 497$ & $1.42 \pm 0.06 \pm 0.11$
\end{tabular}
\caption{{\label{tab:xdssyield}} 
$D_s^{*+}$ yields, efficiency corrected yields and cross sections in the 
specified regions of $x(D_s^{*+})$, where 
${\cal B} \equiv {\cal B}(D_s^{*+}\to D_s^+\gamma){\cal B}
(D_s^+ \to \phi\pi^+){\cal B}(\phi \to K^+K^-)$.
The uncertainty in the efficiency corrected yields and cross-sections 
includes both statistical and systematic error. When two errors are 
presented, the first is statistical while the second is systematic. 
}
\end{table}

\begin{table}
\centering
\begin{tabular}{ c c c c }
                  & Measured      & Efficiency Corrected & \\
$x(D_s^+)$ region & $D_s^+$ Yield & $D_s^+$ Yield  & 
${\cal B}\cdot\sigma(D_s^+)$(pb)\\ \hline
0.44 - 0.50 & $546 \pm 43$  & $1491 \pm 140$ & $0.32 \pm 0.03$\\
0.50 - 0.56 & $663 \pm 40$  & $1795 \pm 139$ & $0.38 \pm 0.03$\\
0.56 - 0.62 & $933 \pm 40$  & $2504 \pm 160$ & $0.53 \pm 0.03$\\
0.62 - 0.68 & $1019 \pm 38$ & $2712 \pm 160$ & $0.57 \pm 0.03$\\
0.68 - 0.74 & $1080 \pm 40$ & $2847 \pm 166$ & $0.60 \pm 0.04$\\
0.74 - 0.80 & $925 \pm 36$  & $2418 \pm 145$ & $0.51 \pm 0.03$\\
0.80 - 0.86 & $759 \pm 31$  & $1968 \pm 122$ & $0.42 \pm 0.03$\\
0.86 - 0.92 & $404 \pm 24$  & $1038 \pm 79$  & $0.22 \pm 0.02$\\
0.92 - 0.98 & $170 \pm 14$  & $433 \pm 42$   & $0.09 \pm 0.01$\\
(0.92 - 1.00) & ($187 \pm 17$) & ($476 \pm 50$) & ($0.10 \pm 0.01$) \\ \hline
0.44 - 1.00 & $6516 \pm 106$ & $17250 \pm 281 \pm 528$ & 
$3.65 \pm 0.06 \pm 0.11$        
\end{tabular}
\caption{{\label{tab:xdsyield1}}
$D_s^+$ yields and cross-sections in the specified regions of $x(D_s^+)$, 
where 
${\cal B} \equiv {\cal B}(D_s^+ \to \phi\pi^+){\cal B}(\phi \to K^+K^-)$.
The $D_s^+$ yields are efficiency corrected using smoothed efficiency values.
The uncertainty in the efficiency corrected yields and cross-sections includes
both statistical and systematic error. When two errors are presented, the 
first is statistical while the second is systematic.
}
\end{table}

\begin{table}
\centering
\begin{tabular}{ c c c c }
                  & Measured          &Efficiency Corrected &  \\ 
$x(D_s^+)$ region & $D_s^{*+}$ Yield  & $D_s^{*+}$ Yield & 
${\cal B}\cdot\sigma(D_s^{*+})$(pb) \\ \hline
0.44 - 0.50 & $104 \pm 21$   & $577 \pm 129$   & $0.13 \pm 0.03$\\
0.50 - 0.56 & $145 \pm 19$   & $808 \pm 133$   & $0.18 \pm 0.03$\\
0.56 - 0.62 & $200 \pm 20$   & $1122 \pm 152$  & $0.25 \pm 0.03$\\
0.62 - 0.68 & $181 \pm 18$   & $1020 \pm 138$  & $0.23 \pm 0.03$\\
0.68 - 0.74 & $208 \pm 18$   & $1180 \pm 148$  & $0.27 \pm 0.03$\\
0.74 - 0.80 & $182 \pm 16$   & $1040 \pm 133$  & $0.23 \pm 0.03$\\
0.80 - 0.86 & $149 \pm 14$   & $852 \pm 113$   & $0.19 \pm 0.03$\\
0.86 - 0.92 & $70 \pm 9$     & $501 \pm 202$   & $0.11 \pm 0.05$\\
0.92 - 0.98 & $16 \pm 5$     & $63 \pm 39$     & $0.01 \pm 0.01$\\
(0.92 - 1.00) & ($15 \pm 5$) & ($60 \pm 37$)   & ($0.01 \pm 0.01$)\\ \hline
0.44 - 1.00 & $1253 \pm 49$  & $7160 \pm 279 \pm 550$ & 
$1.61 \pm 0.06 \pm 0.11$   
\end{tabular}
\caption{{\label{tab:xdsyield2}}
$D_s^{*+}$ yields and cross-sections in the specified regions of $x(D_s^+)$, 
where 
${\cal B} \equiv {\cal B}(D_s^+ \to \phi\pi^+){\cal B}(\phi \to K^+K^-)$. 
The first seven $D_s^{*+}$ yields are efficiency corrected using the smoothed 
efficiency values while the efficiencies for $x(D_s^+)>0.86$ are corrected 
using the raw efficiency values. The uncertainty in the efficiency corrected
yields and cross-sections includes both statistical and systematic error. 
When two errors are presented the first is statistical while the second 
systematic.
}
\end{table}

\begin{table}
\centering
\begin{tabular}{ c c c}
$x(D_s^+)$ region & Primary $D_s^+$ Yield & Primary 
${\cal B}\cdot\sigma(D_s^+)$(pb)\\ \hline
0.44 - 0.50 & $878 \pm 197$  & $0.19 \pm 0.04$\\
0.50 - 0.56 & $937 \pm 200$  & $0.20 \pm 0.04$\\
0.56 - 0.62 & $1313 \pm 230$ & $0.28 \pm 0.05$\\
0.62 - 0.68 & $1630 \pm 219$ & $0.35 \pm 0.05$\\
0.68 - 0.74 & $1594 \pm 231$ & $0.34 \pm 0.05$\\
0.74 - 0.80 & $1315 \pm 204$ & $0.28 \pm 0.04$\\
0.80 - 0.86 & $1063 \pm 173$ & $0.23 \pm 0.04$\\
0.86 - 0.92 & $506 \pm 229$  & $0.11 \pm 0.05$\\
0.92 - 0.98 & $370 \pm 57$   & $0.08 \pm 0.01$\\
(0.92 - 1.00) & ($417 \pm 64)$ & ($0.09 \pm 0.01$)\\ \hline
0.44 - 1.00 & $9652 \pm 408 \pm 760$ & $2.05 \pm 0.09 \pm 0.16$
\end{tabular}
\caption{{\label{tab:dsprim}}
Calculated primary $D_s^+$ yields and cross sections in the specified regions 
of $x(D_s^+)$. The cross-section is presented as ${\cal B} \cdot \sigma$ where 
${\cal B} \equiv {\cal B}(D_s^+ \to \phi\pi^+){\cal B}(\phi \to K^+K^-)$. 
The errors in each $x(D_s^+)$ bin include both statistical and systematic 
uncertainty. When two errors are presented, the first is statistical while 
the second is systematic. 
}
\end{table}

By summing the efficiency corrected $D_s^{*+}$ and primary $D_s^+$ yields 
listed in Tables \ref{tab:xdssyield} and \ref{tab:dsprim}, respectively,  
Eq.~(\ref{eq:pv1}) could be used to calculate $P_V$ for $x(D_s^{(*)+})>0.5$. 
However, the uncertainties in the $D_s^{*+}$ yields are essentially counted 
twice due to the subtraction used to calculate the primary $D_s^+$ yields.
$P_V$ can however be calculated in a way that avoids this subtraction. Since 
all observed $D_s^{*+}$ mesons are assumed to be primary, all observed 
$D_s^+$ mesons are assumed to be either primary or $D_s^{*+}$ daughters, and 
all $D_s^{*+}$ are expected to decay to a $D_s^+$, Eq.~(\ref{eq:pv1}) can be 
rewritten as 
\begin{equation}
P_V = \frac{T(D_s^{*+})}{T(D_s^+)}~, \label{eq:pv2}
\end{equation}
where $T(M)$ is the total number of $M$ mesons in the CLEO II data sample. 
In terms of the quantities measured using the decay modes chosen for this 
analysis,
\begin{equation}
P_V = \frac{n(D_s^{*+})}{n(D_s^+){\cal B}(D_s^{*+}\to D_s^+\gamma)}~, 
\label{eq:pv3}
\end{equation}
where $n(M)$ is the efficiency corrected yield of $M$ mesons in a particular 
$x(D_s^+)$ region. Using this method, 
$P_V(x(D_s^+)>0.44){\cal B}(D_s^{*+}\to D_s^+ \gamma)=0.42 \pm 0.02$.
Using the value 
${\cal B}(D_s^{*+}\to D_s^+ \gamma)=(94.2\pm2.5)\%$~\cite{dspi0} leads to
$P_V(x(D_s)>0.44)=0.44 \pm 0.02(stat.) \pm 0.01(br.)$.


\section{Systematic Uncertainty}

The systematic error for the total $D_s^+$ and $D_s^{*+}$ yields is determined
by varying the selection and fitting procedures as described below and taking 
the variance in the total yield as the estimate of the error. The variance is 
also determined on a bin-by-bin basis and the average percentage variance in 
the individual bins is taken as the estimated systematic uncertainty for all 
bins. The uncertainties in the $D_s^{*+}$ yields for the range 
$0.86<x(D_s^+)<1.0$ are averaged separately since those values are not 
smoothed and the errors are quite large due to the limited number of 
$D_s^{*+}$ events in that region. Systematic uncertainties on the various 
yields are listed in Tables \ref{tab:dssyst} and \ref{tab:dsssyst}.

\begin{table}
\centering
\begin{tabular}{c c}
Variation & Percent Variance \\ \hline
$D_s^+$ peak in $M(KK\pi)$ fit with Gaussian & 2\%(3\%) \\
$D^+$ peak in $M(KK\pi)$ fit with double Gaussian & 1\%(1\%) \\
$D_s^+$ background fit with quadratic & 2\%(3\%)
\end{tabular}
\caption{{\label{tab:dssyst}} Percent variance in total(bin-by-bin) $D_s^+$ 
yields compared to the nominal yields due to the listed sources of systematic 
error. 
}
\end{table}

\begin{table}
\centering
\begin{tabular}{c c}
Variation & Percent Variance \\ \hline
15 MeV/$c^2$ wide $M(KK\pi)$ signal region & 1\%(3\%)\\
25 MeV/$c^2$ wide $M(KK\pi)$ signal region & 1\%(2\%)\\
$\Delta M$ peak fit with Gaussian & 1\%(1\%) \\
$\Delta M$ peak fit with double bifurcated Gaussian & 2\%(3\%) \\
$\cos \theta_s<0.6$ & 6\%(6\%) \\
$E(\gamma)>90,110$ MeV & 2\%(3\%) \\
Uncertainty in $\gamma$ efficiency from ${\cal B}
(\eta \to \gamma\gamma)/{\cal B}(\eta \to 3\pi^0)$ study & 3\%(3\%)
\end{tabular}
\caption{{\label{tab:dsssyst}}Percent variance in total(bin-by-bin) 
$D_s^{*+}$ yields compared to the nominal yields due to the listed sources 
of systematic error. 
}
\end{table}

The acceptance angles for showers implicitly alter the acceptance of tracks 
since there is a high degree of correlation between the flight directions of 
the $D_s^+$ and the photon in the detector. There is also a correlation 
between the photon energy and the decay angle of the $D_s^{*+}$. Varying the 
shower acceptance angles to $|\cos \theta_s|<0.5$ changes the total $D_s^{*+}$
yields and the bin-by-bin yields by approximately 6\%, while changing the 
minimum shower energy to either 90 MeV or 110 MeV changes the total $D_s^{*+}$
yield by approximately 2\% and the bin-by-bin yields by approximately 3\%. 
A 3\% overall systematic uncertainty in photon reconstruction has been 
estimated by comparing the world average value of 
${\cal B}(\eta \to \gamma\gamma)/{\cal B}(\eta-3\pi^0)$~\cite{pdg} with
the relative yields of $\eta \to \gamma \gamma$ and $\eta \to 3\pi^0$ in data 
and Monte Carlo. 

Additional uncertainty exists because of differences in invariant mass 
distributions between data and Monte Carlo and possible inadequacies of the 
fitting functions used to determine the yields. This uncertainty is estimated 
by altering the fitting shapes used to obtain the $D_s^+$ and $D_s^{*+}$ 
yields. Varying the fitting technique for the $M(KK\pi)$ projections by e.g. 
using a Gaussian for the $D_s^+$ signal peak, a double Gaussian with common 
mean for the $D^+$ signal peak, or a second-order polynomial for the 
background alters the total $D_s^+$ yield by approximately 3\% and the 
bin-by-bin yields by approximately 4\%. Using a single Gaussian or double 
bifurcated Gaussian with a common mean for the peak in the $\Delta M$ 
distribution alters the total $D_s^{*+}$ yield by approximately 2\% and the 
bin-by-bin yields by approximately 3\%. 

There is also an uncertainty related to the requirement that $M(KK\pi)$ be 
within 20 MeV/$c^2$ of its nominal value. Widening this requirement to 
25 MeV/$c^2$ and narrowing it to 15 MeV/$c^2$ has resulted in an approximate 
2\% error in the total $D_s^{*+}$ yield and an approximate 4\% error in 
the bin-by-bin yields.

The uncertainties in the efficiency values shown in Tables \ref{tab:dsseff} 
and \ref{tab:dseff} vary for each region of $x(D_s^+)$ and $x(D_s^{*+})$ due 
to limited Monte Carlo statistics and the smoothing process. For instance, 
the errors in the smoothed efficiency values near the limits of the $x$ region
studied are higher than those in the middle of the region due to the 
uncertainty in the slope of the function used in the smoothing process. The 
errors in the efficiency contribute to the systematic uncertainty on a 
bin-by-bin basis and the percentage errors are added in quadrature for the
determination of the percentage error for the total yields.

All of the individual systematic uncertainties associated with a given yield 
are added together in quadrature with the percentage error in the efficiency 
to determine the total systematic uncertainty in the $D_s^+$ and $D_s^{*+}$ 
yields. These systematic uncertainties are already included in the errors in 
the yields in Tables \ref{tab:xdssyield}, \ref{tab:xdsyield1}, 
\ref{tab:xdsyield2} and \ref{tab:dsprim}. After including the total systematic
uncertainty,
$P_V(x(D_s^+)>0.44)=0.44 \pm 0.02(stat.) \pm 0.03(syst.) \pm 0.01(br.)$.


\section{Discussion of Results}

The momentum distributions of hadrons created in the fragmentation process are
commonly modeled with either the Andersson {\it et al.} symmetric 
fragmentation function~\cite{andersson} or the Peterson {\it et al.} 
fragmentation function~\cite{peterson}. Both of these functions depend upon
$z=\frac{E_h+p_\parallel}{E+p}$, where $E_h$ is the energy of the hadron, 
$p_\parallel$ is the hadron momentum parallel to $p$, the momentum of the 
primary quark from the production process, and $E$ is the energy of the 
primary quark. The Andersson function is 
\begin{equation}
f(z) \propto z^{-1}(1-z)^a\exp(-b\ m_\bot^2/z)~, \label{eq:andersson}
\end{equation}
where $a$ and $b$ are free parameters, $m_\bot=\sqrt{m_q^2+p_\bot^2}$, 
$m_q$ is the mass of the primary quark and $p_\bot$ is the hadron momentum 
perpendicular to $p$. The Peterson function is
\begin{equation}
f(z) \propto \frac{1}{z[1-(1/z)-\epsilon_P/(1-z)]^2}~, \label{eq:peterson}
\end{equation}
where $\epsilon_P$ is the single free parameter.

To properly compare fragmentation models with data it is necessary to use the 
above functions in a full Monte Carlo simulation that incorporates photon 
radiation, gluon radiation and other effects. To facilitate comparison with 
other experimental results, $x$ is used as an approximation of $z$ and a 
binned $\chi^2$ fit to the data is performed using these two functions as 
shown in Figs. \ref{fig:dssqq2} and \ref{fig:dsqq2}. Since the parameters 
$b$ and $m_\bot$ only appear in Eq. (\ref{eq:andersson}) as a product, the 
constraint $b=1$ has been used for the fit, thereby changing the 
interpretation of the value of $m_\bot$. The numerical results from the fits 
are listed in Table \ref{tab:fits}. The normalizations of these fits are not 
used to calculate a value of $P_V$ due to differences between $x$ and $z$ 
that are non-negligible in the low $D_s$ momentum regime.

\begin{figure}
\centerline{\psfig{file=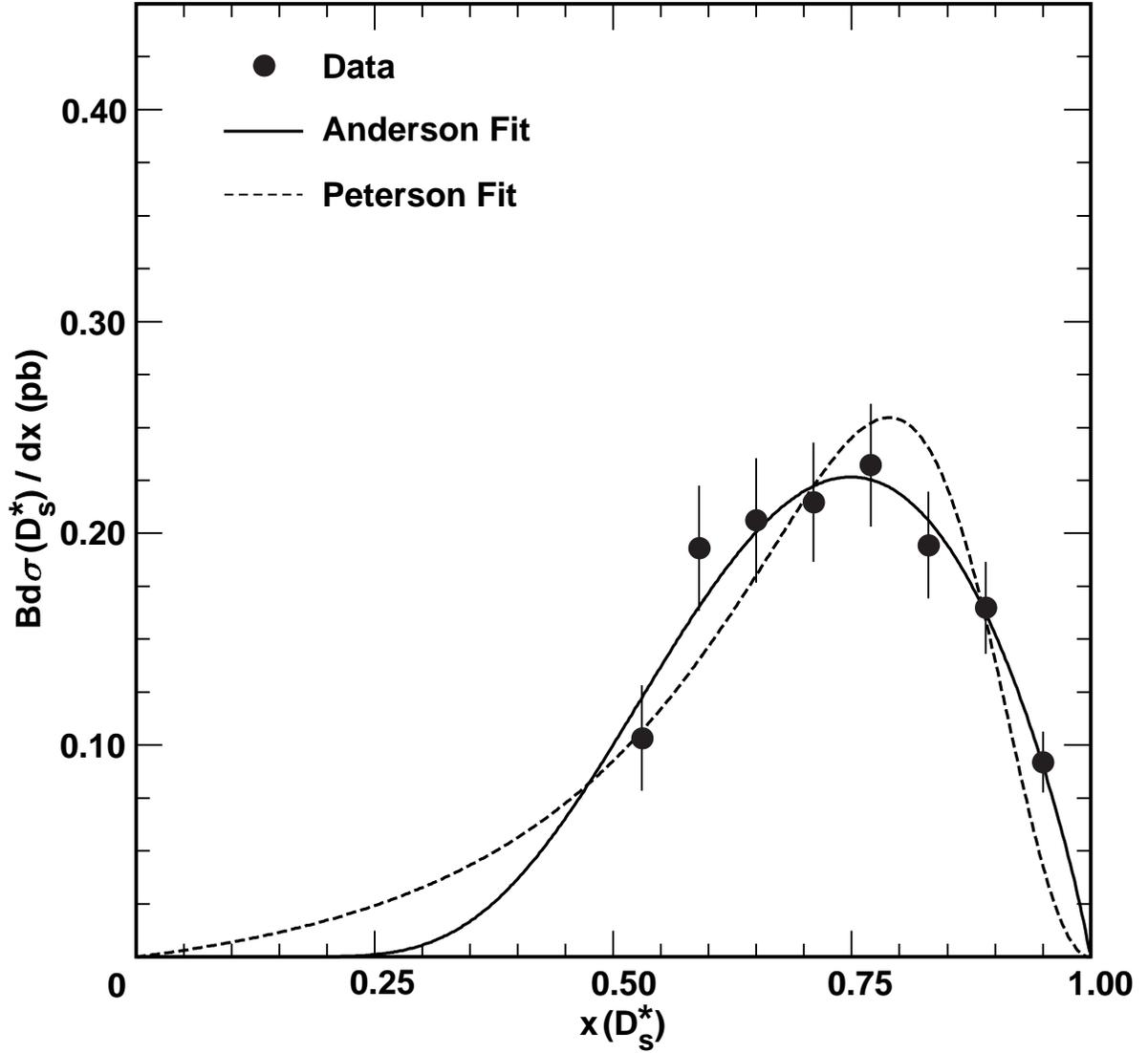,height=6in}}
\caption{{\label{fig:dssqq2}}
${\cal B}\cdot \sigma(D_s^{*+})$ spectrum fit with the Andersson {\it et al.} 
and Peterson {\it et al.} fragmentation functions, where 
${\cal B}\equiv{\cal B}(D_s^{*+}\to D_s^+\gamma )
{\cal B}(D_s^+ \to \phi \pi^+){\cal B}(\phi \to K^+K^-)$.
}
\end{figure}

\begin{figure}
\centerline{\psfig{file=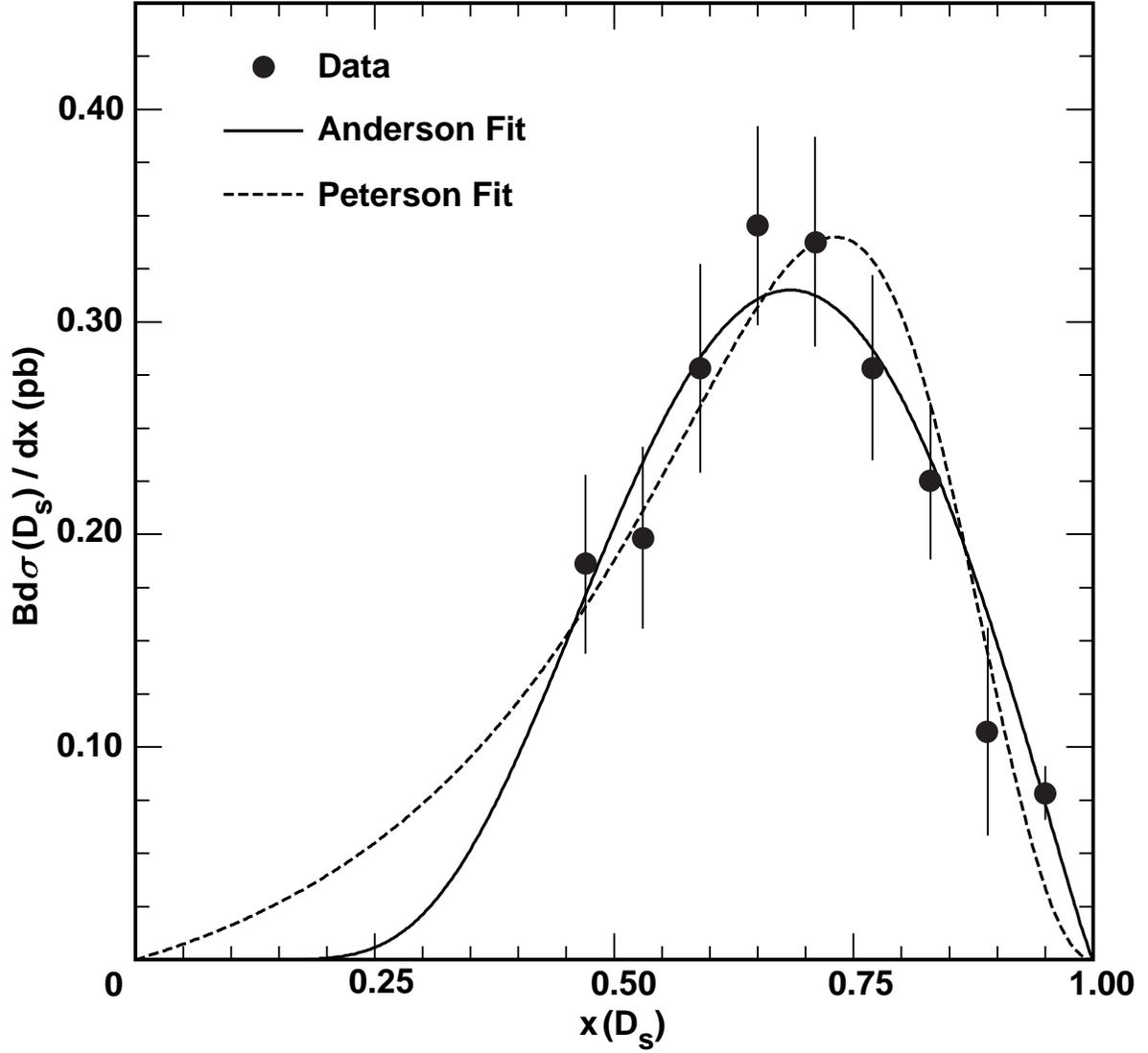,height=6in}}
\caption{{\label{fig:dsqq2}}
Primary ${\cal B}\cdot\sigma(D_s^+)$ spectrum fit with the Andersson 
{\it et al.} and Peterson {\it et al.} fragmentation functions, where 
${\cal B}\equiv{\cal B}(D_s^+ \to \phi \pi^+){\cal B}(\phi \to K^+K^-)$.
}
\end{figure}

\begin{table}
\centering
\begin{tabular}{ l l c} 
 & \multicolumn{1}{c}{Fit Results} & $\chi^2$/d.o.f. \\ \hline
Andersson: & & \\
$D_s^{*+}$:& $a=0.9\pm0.2$, $m_\bot=1.7\pm0.1$ & 1.9/5 \\
$D_s^+$:& $a=1.1\pm0.2$, $m_\bot=1.5\pm0.1$ & 3.2/6 \\
Peterson: & & \\
$D_s^{*+}$:& $\epsilon_P=0.056\pm0.008$ & 20.5/6 \\
$D_s^+$:& $\epsilon_P=0.10\pm0.02$ & 17.4/7
\end{tabular}
\caption{{\label{tab:fits}} 
Results of fits to the $D_s^{*+}$ and $D_s^+$ spectra in $x(D_s^{*+})$ and 
$x(D_s^+)$, respectively, with the Anderson {\it et al.} and Peterson 
{\it et al.} analytical fragmentation functions.
}
\end{table}

The fragmentation spectra for charm mesons has been studied previously by the 
CLEO collaboration\cite{cleo88} and input parameters for the Andersson 
{\it et al.} model were determined using measured fragmentation distributions 
for $D^{*+}$, $D^0$, $D^+$, $D_s$ and $\Lambda_c$. A comparison of the data 
presented here with a Monte Carlo distribution using the parameters determined
in that study, $a=0.60$ and $b=0.52$, is shown in Figure \ref{fig:qqall}. 
The use of these values as input parameters for charm fragmentation into 
$D_s$ mesons clearly result in a momentum distribution that is too soft, 
which is not surprising since P-wave charm meson decays to $D^{*+}$, $D^0$ 
and $D^+$ were not excluded in the prior study.  

\begin{figure}
\centerline{\psfig{file=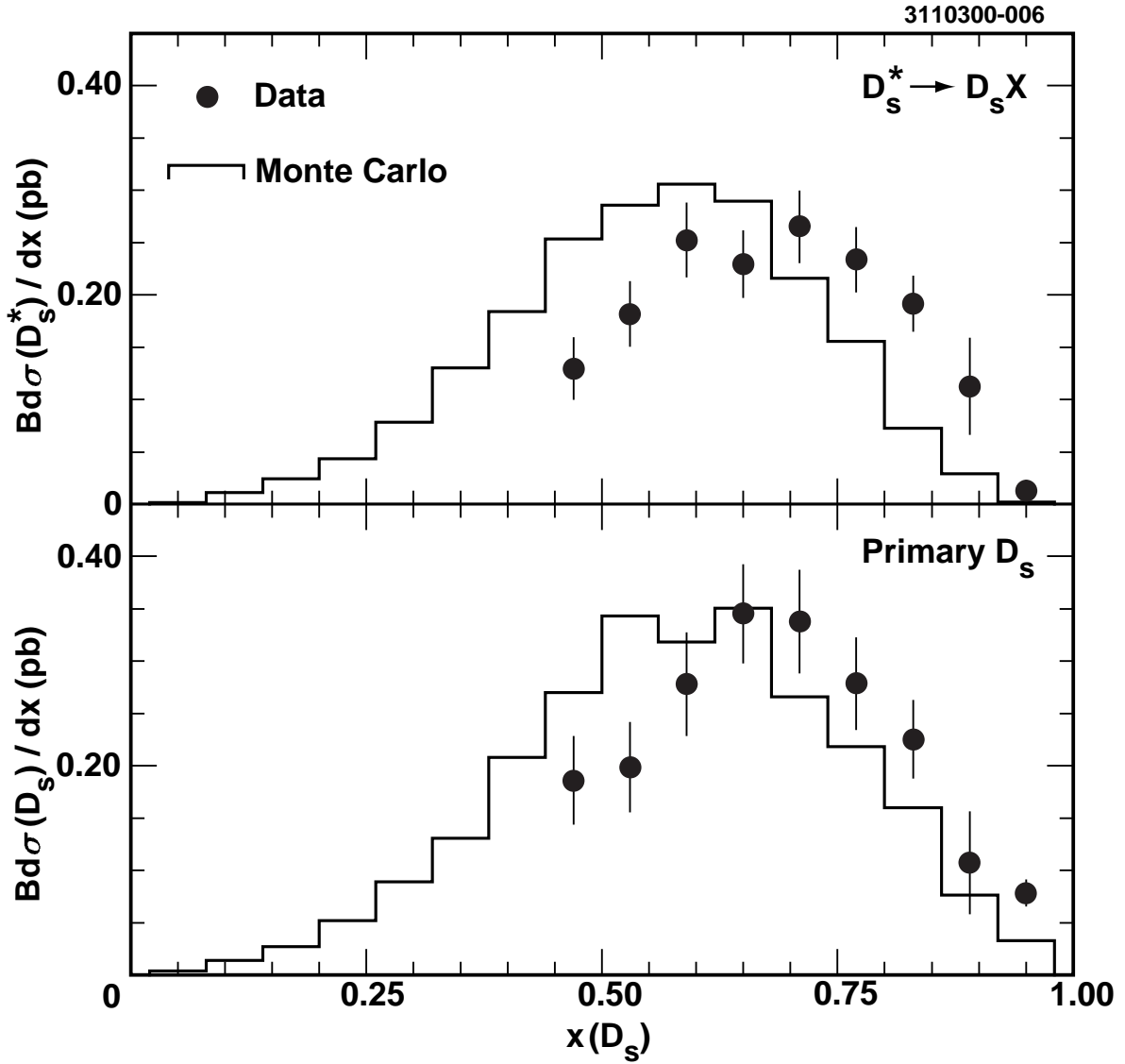,height=6in}}
\caption{{\label{fig:qqall}}
${\cal B}\cdot\sigma(D_s^{*+})$ and ${\cal B}\cdot\sigma(D_s^+)$ for primary 
$D_s$ compared to the Monte Carlo production spectrum using the Andersson 
{\it et al.} fragmentation function parameters $a=0.60$ and $b=0.52$~[9], 
where 
${\cal B}\equiv {\cal B}(D_s^+ \to \phi \pi^+){\cal B}(\phi \to K^+K^-)$.
}
\end{figure}

High levels of combinatoric background at low values of $x(D_s^+)$ prohibit 
a good measurement of $P_V$ for the full range of allowed $D_s^+$ momenta.
Based on Monte Carlo simulations and the data presented, approximately 
$75-85\%$ of all $D_s^{*+}$ and $D_s^+$ are expected to have $x(D_s^+)>0.44$ 
and the value of $P_V$ presented here is not expected to differ much from 
$P_V$(all $x$).

It is possible to make a model-dependent extrapolation of $P_V$(all $x$)
using 
\begin{equation}
P_V(x>0)=\frac{1}{1+\left (\frac{1}{P_V(x>0.44)}-1\right )\frac{Q_V}{Q_P}}~,
\label{eq:extrapolate}
\end{equation}
where $Q_V$ is the percentage of $D_s^{*+}$ that decay to a $D_s^+$ with
$x(D_s^+)>0.44$ and $Q_P$ is the fraction of primary $D_s^+$ that have 
$x(D_s^+)>0.44$.  Since only about one fifth of either fragmentation spectra 
lies below $x(D_s^+)=0.44$, and because both distributions approach zero 
smoothly as $x(D_s^+) \to 0$, the ratio $Q_V/Q_P$ is expected to be close to 
unity and to only depend weakly upon the chosen fragmentation parameters.

Using the Andersson {\it et al.} model with the parameters $a=0.60$ and
$b=0.52$ results in $Q_V=0.773$, $Q_P=0.795$, $Q_V/Q_P=0.972$, and from Eq.
(\ref{eq:extrapolate}), $P_V$(all $x(D_s^+))=0.45 \pm 0.05$. Changing the 
input parameters to provide a harder spectrum has a very small effect on 
$Q_V/Q_P$.  A distribution created with $a=0.4$ and $b=0.9$, for example, 
provides a much improved representation of the data and results in 
$Q_V=0.860$, $Q_P=0.875$, $Q_V/Q_P=0.983$ and $P_V$(all $x(D_s^+))=0.45 \pm
0.05$. This clearly shows that the dependence of the $P_V$ extrapolation on 
the choice of fragmentation parameters is indeed weak. 

Based on the results of varying the input parameters for the two models, a
systematic uncertainty of 3\% is estimated for the model-dependent
extrapolation resulting in a final extrapolated value of 
$P_V$(all $x(D_s^+))=0.45 \pm 0.05$, which is significantly different than
the expected result based on spin counting. 

Other measurements of $P_V$ for charm and bottom mesons have been 
presented~\cite{aleph2,opal3,sld,l3,opal2}, but it is difficult
to make direct comparisons between those results and the one presented here 
because of differences in methodology and center-of-mass energies in the 
other analyses. Nonetheless, measurements of $P_V(B)$ are generally close to 
the spin-counting expectation while measurements of $P_V(D)$ are well below 
that value as shown in Table \ref{tab:prev}.

\begin{table}
\centering
\begin{tabular}{c c}
Collaboration & Result \\ \hline
ALEPH~\cite{aleph2} & $P_V(D_s)=0.60 \pm 0.19$\\
ALEPH~\cite{aleph2} & $P_V(D)=0.60 \pm 0.05$  \\
OPAL~\cite{opal3}   & $P_V(D)=0.57 \pm 0.06$  \\
SLD~\cite{sld}      & $P_V(D)=0.57 \pm 0.07$  \\  
L3~\cite{l3}        & $P_V(B)=0.76 \pm 0.10$  \\
OPAL~\cite{opal2}   & $P_V(B)=0.76 \pm 0.09$  
\end{tabular}
\caption{{\label{tab:prev}} 
Results of previous measurements of $P_V$ for heavy quark mesons at other 
experiments. All of these measurements used data samples collected at $Z^0$ 
resonance.}
\end{table}


\section{Conclusion}
In summary, studies of $D_s^{*+}$ and $D_s^+$ fragmentation in $e^+e^-$ 
annihilations at $\sqrt{s}=10.5$ GeV have been presented. $P_V(x(D_s)>0.44)$ 
has been measured to be $0.44 \pm 0.02(stat.) \pm 0.03(syst.) \pm 0.01(br.)$. 
When extrapolated to the entire available momentum region this measurement 
deviates significantly from $P_V=0.75$, the expected result based on simple 
spin counting.  

\section{Acknowledgements}

We gratefully acknowledge the effort of the CESR staff in providing us with
excellent luminosity and running conditions.
J.R. Patterson and I.P.J. Shipsey thank the NYI program of the NSF, 
M. Selen thanks the PFF program of the NSF, 
M. Selen and H. Yamamoto thank the OJI program of DOE, 
J.R. Patterson, K. Honscheid, M. Selen and V. Sharma 
thank the A.P. Sloan Foundation, 
M. Selen and V. Sharma thank the Research Corporation, 
F. Blanc thanks the Swiss National Science Foundation, 
and H. Schwarthoff and E. von Toerne thank 
the Alexander von Humboldt Stiftung for support.  
This work was supported by the National Science Foundation, the
U.S. Department of Energy, and the Natural Sciences and Engineering Research 
Council of Canada.


\end{document}